# Advanced LIGO Two-Stage Twelve-Axis Vibration Isolation and Positioning Platform

## Part 2: Experimental Investigation and Tests Results

**F. Matichard[1,2,*], B. Lantz[3], K. Mason[1], R. Mittleman[1], B. Abbott[2], S. Abbott[2], E. Allwine[5], S. Barnum[1], J. Birch[4], S. Biscans[1], D. Clark[3], D. Coyne[2], D. DeBra[3], R. DeRosa[6], S. Foley[1], P. Fritschel[1], J.A. Giaime[4,6], C. Gray[5], G. Grabeel[5], J. Hanson[4], M. Hillard[1], J. Kissel[5], C. Kucharczyk[3], A. Le Roux[4], V. Lhuillier[5], M. Macinnis[2], B. O'Reilly[4], D. Ottaway[1], H. Paris[5], M. Puma[4], H. Radkins[5], C. Ramet[4], M. Robinson[5], L. Ruet[1], P. Sareen[1], D. Shoemaker[1], A. Stein[1], J. Thomas[4], M. Vargas[4], J. Warner[5].**

**[1] MIT, Cambridge, MA, USA**

**[2] Caltech, Pasadena, CA, USA**

**[3] Stanford University, Stanford, CA, USA**

**[4] LIGO Livingston Observatory, Livingston, LA, USA**

**[5] LIGO Hanford Observatory, Hanford, WA, USA**

**[6] Louisiana State University, Baton Rouge, LA, USA**

* **Corresponding Author:** fabrice@ligo.mit.edu

LIGO Project MIT

MIT NW22-295

185 Albany Street

Cambridge, MA 02139 USA

Phone: +001-617-253-6410

Fax: +001-617-253-7014

**Abstract**

This paper presents the results of the past seven years of experimental investigation and testing done on the two-stage twelve-axis vibration isolation platform for Advanced LIGO gravity waves observatories. This five-ton two-and-half-meter wide system supports more than a 1000 kg of very sensitive equipment. It provides positioning capability and seismic isolation in all directions of translation and rotation. To meet the very stringent requirements of Advanced LIGO, the system must provide more than three orders of magnitude of isolation over a very large bandwidth. It must bring the motion below $10^{-11}\ m/\sqrt{Hz}$ at 1 Hz and $10^{-12}\ m/\sqrt{Hz}$ at 10 Hz. A prototype of this system has been built in 2006. It has been extensively tested and analyzed during the following two years. This paper shows how the experimental results obtained with the prototype were used to engineer the final design. It highlights how the engineering solutions implemented not only improved the isolation performance but also greatly simplified the assembly, testing, and commissioning process. During the past two years, five units have been constructed, tested, installed and commissioned at each of the two LIGO observatories. Five other units are being built for an upcoming third observatory. The test results presented show that the system meets the motion requirements, and reach the sensor noise in the control bandwidth.

## 1 Introduction

Gravity wave observatories use km long interferometers in order to detect strain in space-time produced by astrophysical events [1]-[5]. A very high level of vibration and seismic isolation is required to operate such experiments. Different techniques have been developed and used over the years to reach an adequate level of isolation. They include passive stacks, passive suspensions, inverted pendulums, active inertial control, and low frequency passive isolators [6]-[14]. Combinations of these various techniques are often necessary to reach suitable levels of isolation. Beyond isolation performance, experience has shown that operability and robustness are among the primary requirements for such systems. It is critical that they can be assembled, installed, tested and commissioned in a timely and effective manner. They must be robust to ensure high duty cycle during operation.

Advanced LIGO belongs to the new generation of gravity waves detectors that is currently being built [15]. To meet the very stringent requirements, it includes a sophisticated combination of active platforms and passive suspensions [16]-[20]. This paper summarizes the experimental investigation and tests results of the two-stage twelve-axis seismic isolation platform designed to support Advanced

LIGO core optics. Fifteen units are needed for the Advanced LIGO program (Five for each of the US based observatories of Livingston and Hanford, and five for a third observatory whose location abroad is being studied).

This two-stage platform is an In-vacuum Seismic Isolator (ISI) used in the LIGO vacuum chambers called Basic Symmetric Chambers (BSC). It is referred to as the BSC-ISI system. The concept is based on early work done during the nineties by the group at JILA [21]-[24]. They demonstrated the feasibility and benefits of active seismic isolation systems for low frequency sensitive applications. Passive systems with equivalent performance would require very low natural frequencies (below 100 mHz). The high flexibility inherent in such systems usually complicates the assembly and commissioning process. If well designed, an active system using stiffer springs can ease both the assembly and commissioning steps while providing optimal isolation performance at low frequency.

The results they obtained motivated the construction of a rapid prototype for LIGO applications [25]-[26]. This system was a two-stage platform equipped with commercial inertial sensors. Magnetic actuators were used for the drive. The rapid prototype demonstrated that this concept could operate robustly, which is a crucial requirement for a system aimed at supporting the operations of an observatory.

These promising results led to the construction of a technical demonstrator [27]. This system was a full-scale platform designed to validate the two-stage vibration isolation concept as the baseline approach for Advanced LIGO detectors. Like the rapid prototype, this system was made of two stages in series, imbricated to reduce the volume occupied. Spring blades inspired by GEO suspensions were used to provide the vertical flexibility [2]. Flexure rods were used to provide the horizontal flexibility. Magnetic actuators were used for the drive. A combination of long period seismometers and passive geophones were used to sense the inertial motion of the first stage. Low noise commercial passive geophones were used to sense the inertial motion of the second stage. This demonstrator showed that the active system could operate robustly, reliably and meet isolation requirements.

Based on the results of the technical demonstrator, a prototype of a two-stage platform designed for Advanced LIGO detectors was built in 2006 [28]-[30]. The architecture was based on the technical demonstrator: same types of sensors, actuators and spring components. It featured a base-stage opened in the center to access the inverted (down-facing) optical table of the second stage. All instruments were podded in sealed chambers for the platform to be compatible with LIGO ultra-high vacuum requirements.

Extensive testing was done on this prototype during the next two years at the LIGO-MIT facilities (2006-2008) [31]. Results showed that the necessary isolation could be achieved, but that the internal modes of the structure and its payload would complicate and slow down the commissioning process of Advanced LIGO. The excessively high number of internal resonances and their very low damping ratio led to complicated controllers with low robustness. In order to achieve bandwidth objectives (30 Hz unity gain frequency), control filters based on plant inversion compensation techniques had to be implemented. Such an approach was not suitable for robust operation of Advanced LIGO. Many features and options to speed up the assembly process were also identified during this prototyping period.

The test results of the prototyping run were used to engineer the final design (2009-2010) [32]-[33]. The goal was to design a system suitable for timely assembly, testing and commissioning of the fifteen units needed for Advanced LIGO. The design is presented in the first of two companion papers [35]. This second paper presents the experimental investigation and the test results obtained during the prototyping, development and production phases. The next section of this paper gives an overview of the BSC-ISI platform and the system environment to which it belongs. The third section details how the prototyping results have been used to engineer hardware solutions improving the performance and robustness of the active control loops. The fourth section presents the driven transfer functions. The fifth section summarizes the control scheme and presents examples of control loops. The sixth section shows both transmissibility and absolute motion results.

## 2 System Overview

A CAD representation of the BSC-ISI system is shown in Fig. 1 (a) and a picture of a unit in the assembly area of the LIGO Hanford observatory is shown in Fig. 1 (b). A detailed description of the two stage-system architecture and its sub-assemblies is given in the first part of the two companion papers [35].

The conceptual drawing in Fig. 2 represents the BSC-ISI as it is used at the LIGO observatories. It is mounted on a hydraulic pre-isolator located outside of the vacuum system [36]-[37]. The BSC-ISI is installed in vacuum, on the pre-isolator. It provides two stages of isolation. It supports an optical payload that includes four layers of passive isolation [9]-[19]. A CAD representation of this assembly is shown in Fig. 3.

The following sections provide a detailed characterization of the BSC-ISI platform's response. In some tests, the pre-isolator actuators are used to apply forces on Stage 0 for system identification. The vector of forces applied on Stage 0 is called $\{f_0\}$ in Fig. 2. It is made of the three translational forces along the axis of the Cartesian basis and three torques around those axes. The vector of translation and rotation motions is called $\{x_0\}$. Stage 1 forces and displacements vectors are noted $\{f_1\}$, $\{x_1\}$, and Stage 2 forces and displacements vectors are noted $\{f_2\}$, $\{x_2\}$.

(a)

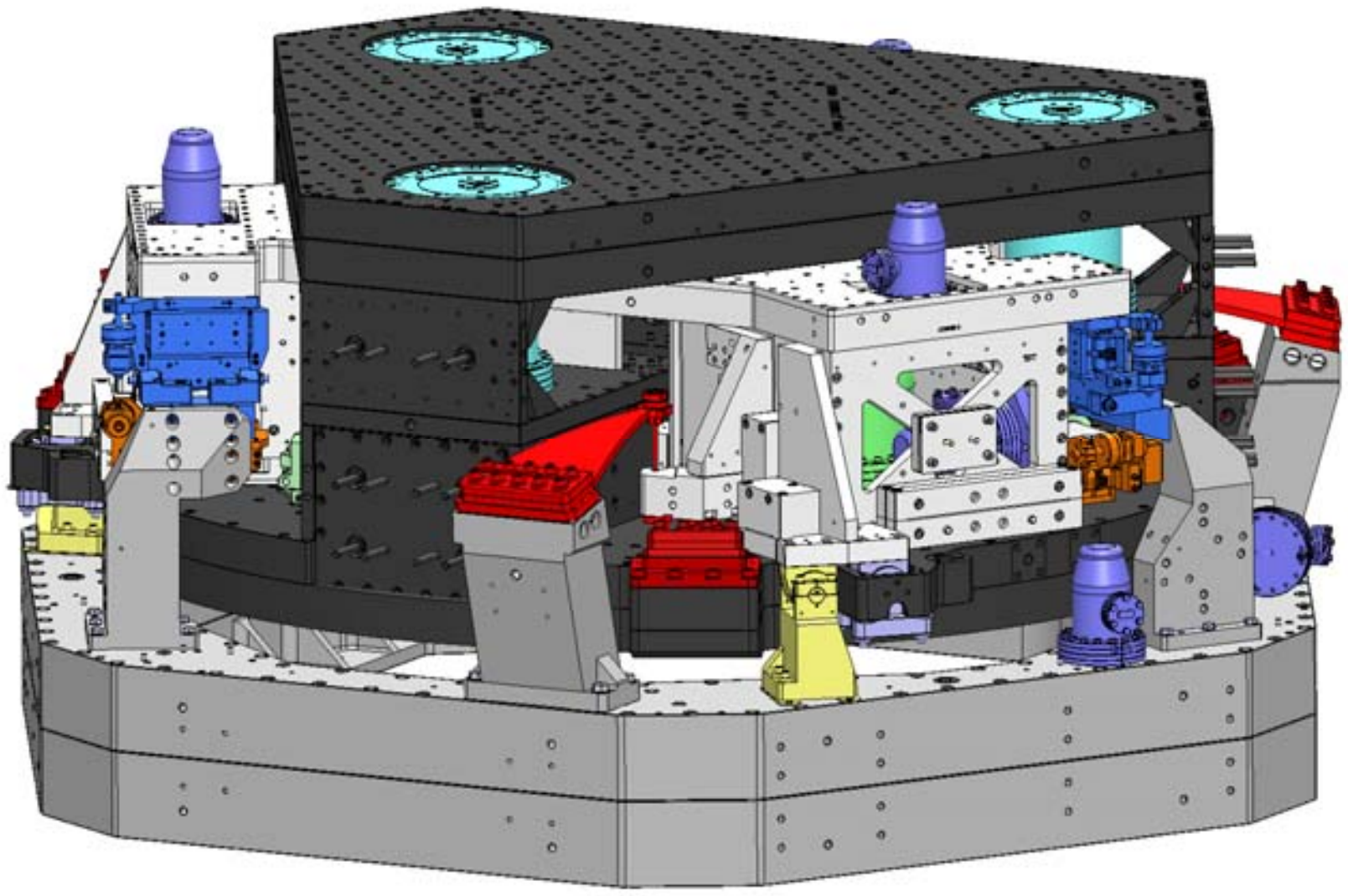

(b)

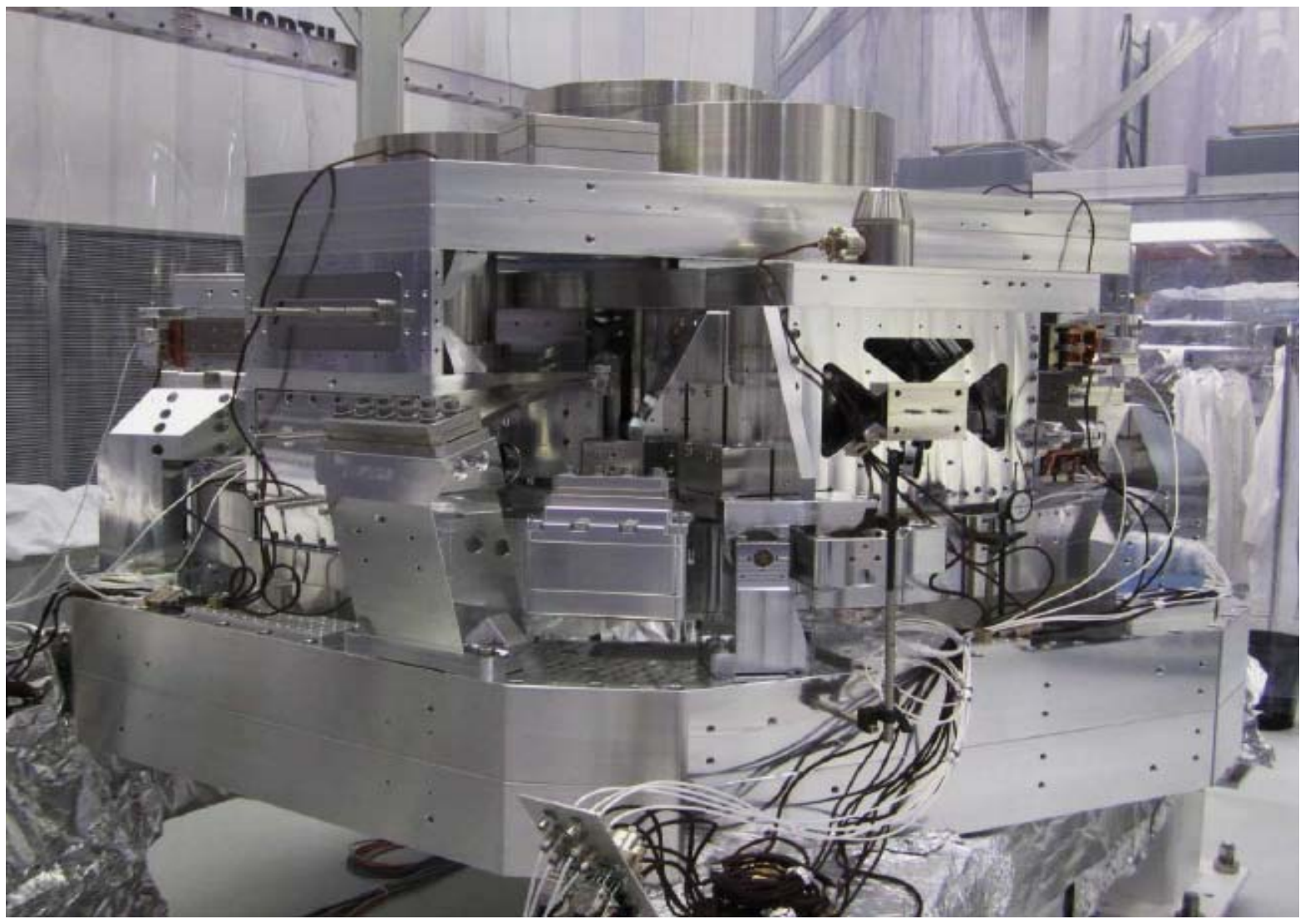

**Fig. 1. (a) CAD representation and picture of a BSC-ISI system. (b) A unit on a test stand at the LIGO Hanford Observatory.**

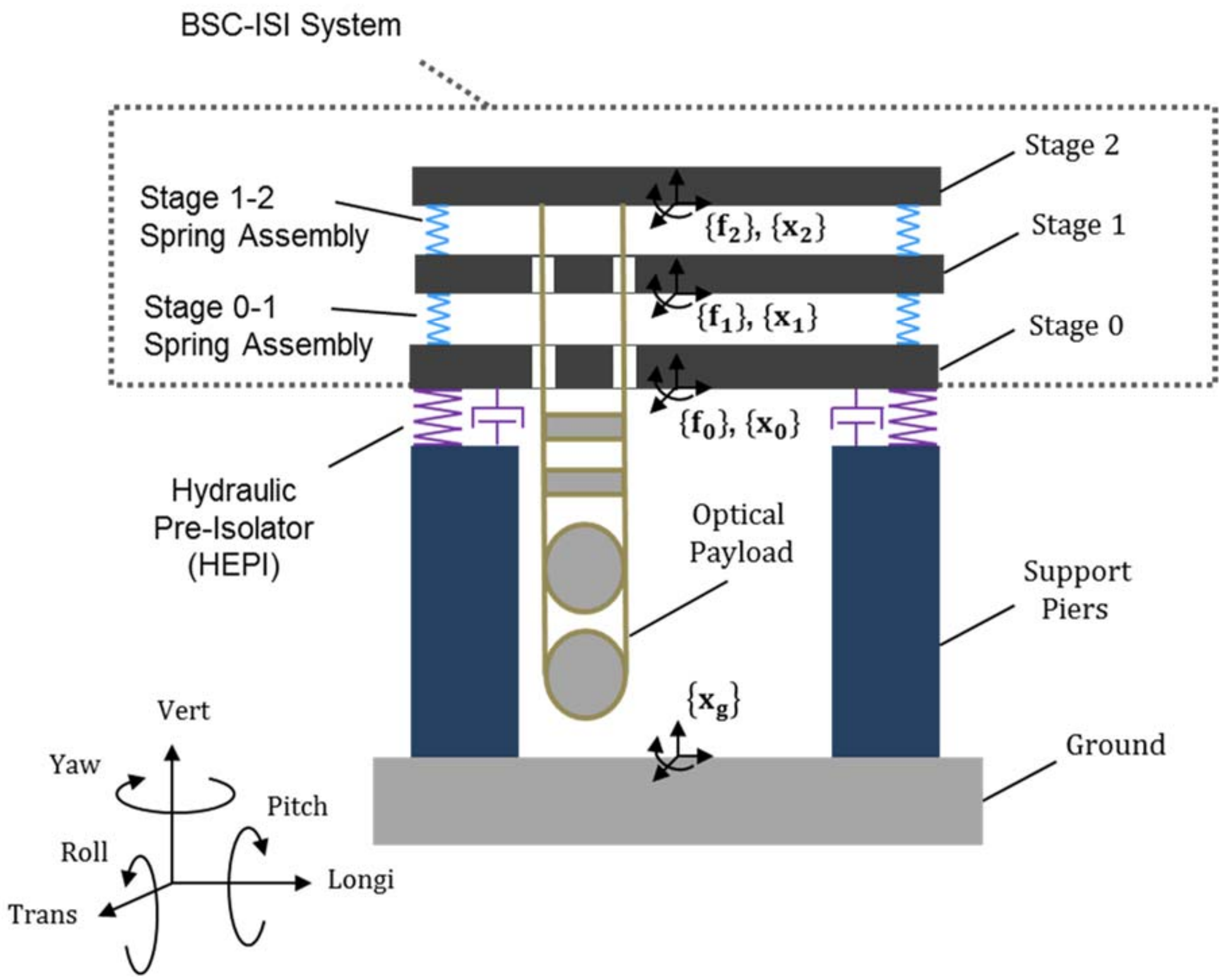

**Fig. 2. Schematic representation of a BSC-ISI platform in the Advanced LIGO system environment.**

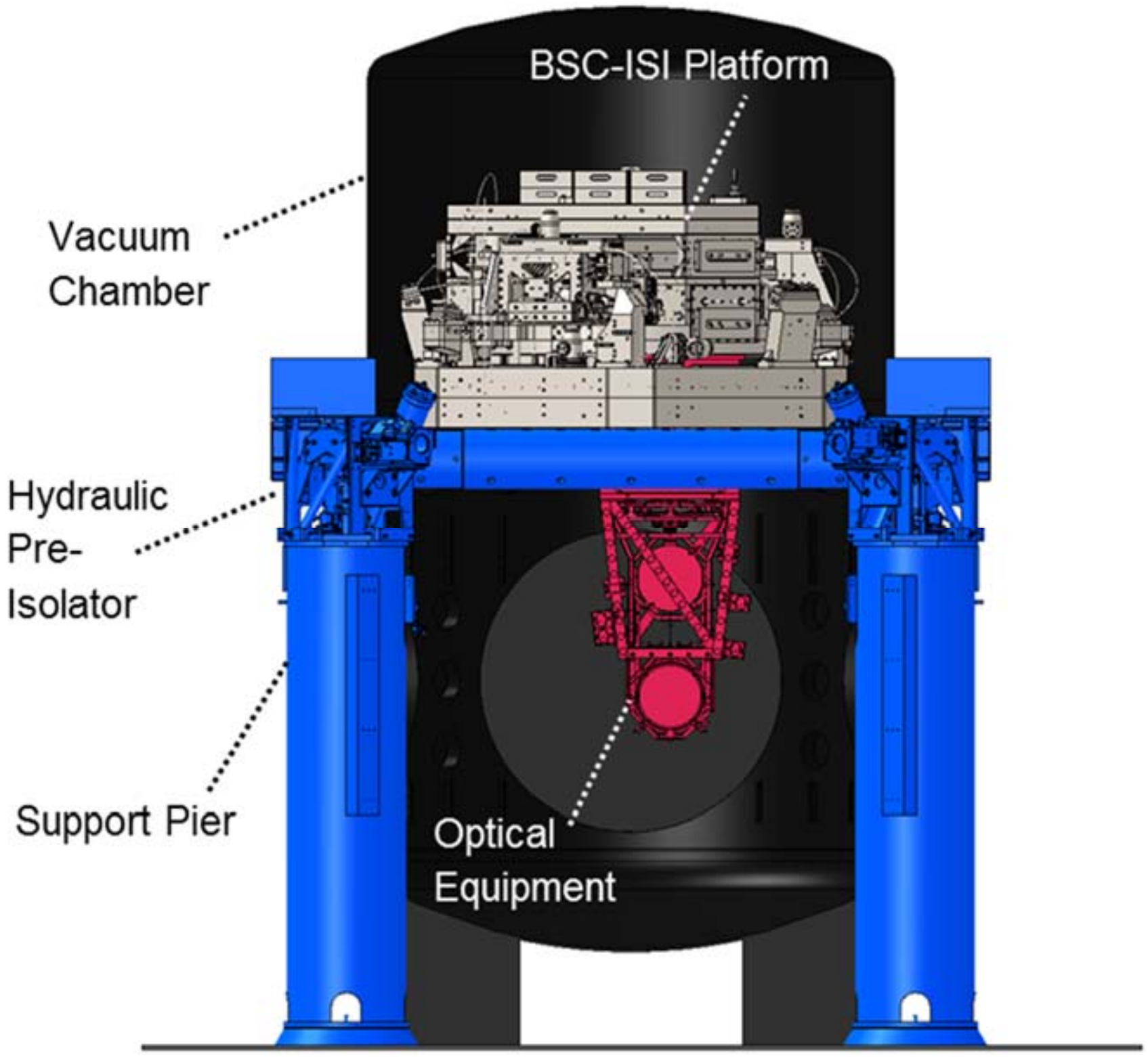

**Fig. 3. CAD representation of a BSC-ISI platform in the Advanced LIGO system environment.**

# 3 Structural analysis and testing

The servo bandwidth and performance of a vibration isolation system are directly related to its higher order dynamics. Rigid-body modes necessary to provide isolation must be greatly separated (in the frequency domain) from the deformation modes. The stiffer the structure, the higher the structural resonance frequencies, and the easier it is to implement the control. This section summarizes how the BSC-ISI structure has been engineered to optimize the active control performance and robustness.

## *3.1 Main structure*

A BSC-ISI prototype was built in 2006 [31]. The feedback control bandwidth goal was to set the unity gain frequency near 30 Hz with at least 35 degrees of phase margin, and 20 dB of gain margin. The stages were designed so that the lowest structural resonances would be above 150 Hz. The dashed curve in Fig. 4 shows an example of a transfer function obtained with the prototype (Plant). A number of local resonances were dominating the system's response at low frequencies. Most of these local resonances were associated with equipment and ballast mounted on the platform. The dash-dotted curve shows a controller that was designed to achieve a 20 Hz unity gain frequency. In order to recover sufficient phase margin in the control bandwidth, the plant response had to be almost completely inverted. The controller has very high-Q features, indicating poor robustness. The open loop shown by the solid curve has little gain margin.

Finite element models and experimental modal analysis were used to identify the local and global modes causing these resonances. Based on these results, the system was re-engineered between 2009 and 2010 to improve the system's dynamics, and consequently the active control performance and robustness [32]-[33].

To this end, some of the initial design requirements were relaxed. For example, the requirement on the distance between the center of mass location and the horizontal actuators plane has been redefined. This requirement is necessary to reduce tilt-horizontal coupling effects at the rigid-body resonances of the open-loop response. Experimental results showed that this offset could be increased without significantly affecting the closed-loop cross couplings. This allowed us to reduce the amount of ballast mass needed to align the center of mass with the actuators, and raising the platform's natural frequencies.

The inertial sensors were also relocated with respect to the actuators. It is usually good practice to collocate sensors and actuators to minimize the phase loss in the open-loop transfer functions and therefore to facilitate the design of the control loops. For this system, maintaining perfect colocation was severely constraining the design. Firstly, the instruments were not located in strategically stiff locations. Therefore they were sensitive to local modes, and close to the maximum displacement of the main structural modes. Secondly, the inertial sensors were sensitive to the actuators' magnetic fields. Experimental results prove this approach to be an excellent compromise.

Numerous design and FEA iterations have been done to increase not only the stages' global structural stiffness but also the local stiffness in the vicinity of the instrumentation [33]-[34]. The preload in the joints of the bolted assembly has also been significantly increased. Comparison between FEA and experimental results showed that the actual bolted structure (experiment) behaved nearly identically to a theoretical monolithic structure (FEA, continuous joints). Comparison of regular and ultra-clean assembly (all components cleaned in chemical bath to dissolve contaminants, and baked to reduce the water content) also showed little reduction of the stages stiffness.

Fig. 5 shows FEA results of modal analysis for the Stage 1 structure free of boundary conditions. Fig. 5 (a) shows the lowest mode obtained with the Stage 1 prototype, and Fig. 5 (b) shows the lowest mode obtained with the Stage 1 of the re-engineered system. The lowest frequency mode has been raised from 150 Hz to 255 Hz. The sensors are re-positioned near the nodes of these low frequency modes.

An experimental setup used to verify these results is shown in Fig. 6. An impact hammer, accelerometer and spectrum analyzer are used to perform the modal analysis. The lowest structural resonance has been measured at 260 Hz, in good agreement with the finite element analysis result. When fully instrumented and connected to the other stages, the lowest resonance remains above 200 Hz. (Typically around 220 Hz, with no more than a couple Hertz of variability from unit to unit).

Fig. 7 shows example of plant and control loop transfer functions for an Advanced LIGO unit (final design). The system's transfer function is shown by the dashed curve (Plant), with a first resonance at 220 Hz. Above that frequency, very few resonances are visible. Mass dampers were designed and installed on all units to damp the first mode. The response of all of the 15 units are close enough that the dampers can be installed interchangeably. They reduce the Q-factor of the main resonance by a factor of 7. The dash-dotted curve shows the controller used to achieve a 25 Hz upper unity gain frequency with 35 degrees of phase margin. Only minor adjustments need to be done to tune the controllers for other units. The open loop curve shows that the gain margin has significantly been improved by comparison with the prototype. An upper unity gain frequency of 40Hz can be obtained with a slightly more complex controller. Quasi-generic controllers can be used to control all the Advanced LIGO BSC-ISI units, thus significantly reducing the commissioning time.

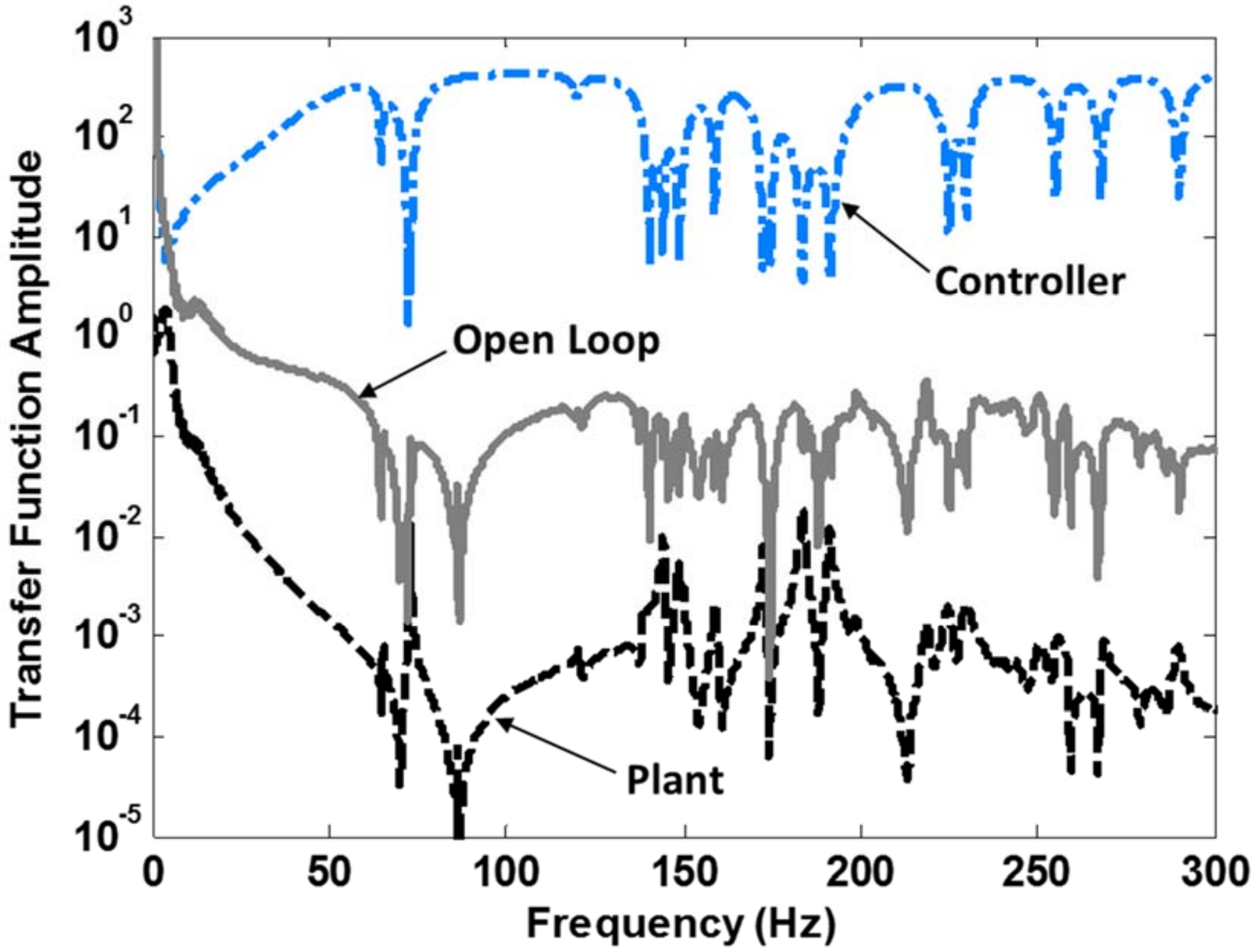

**Fig. 4. Prototype's plant, controller and open loop transfer function.**

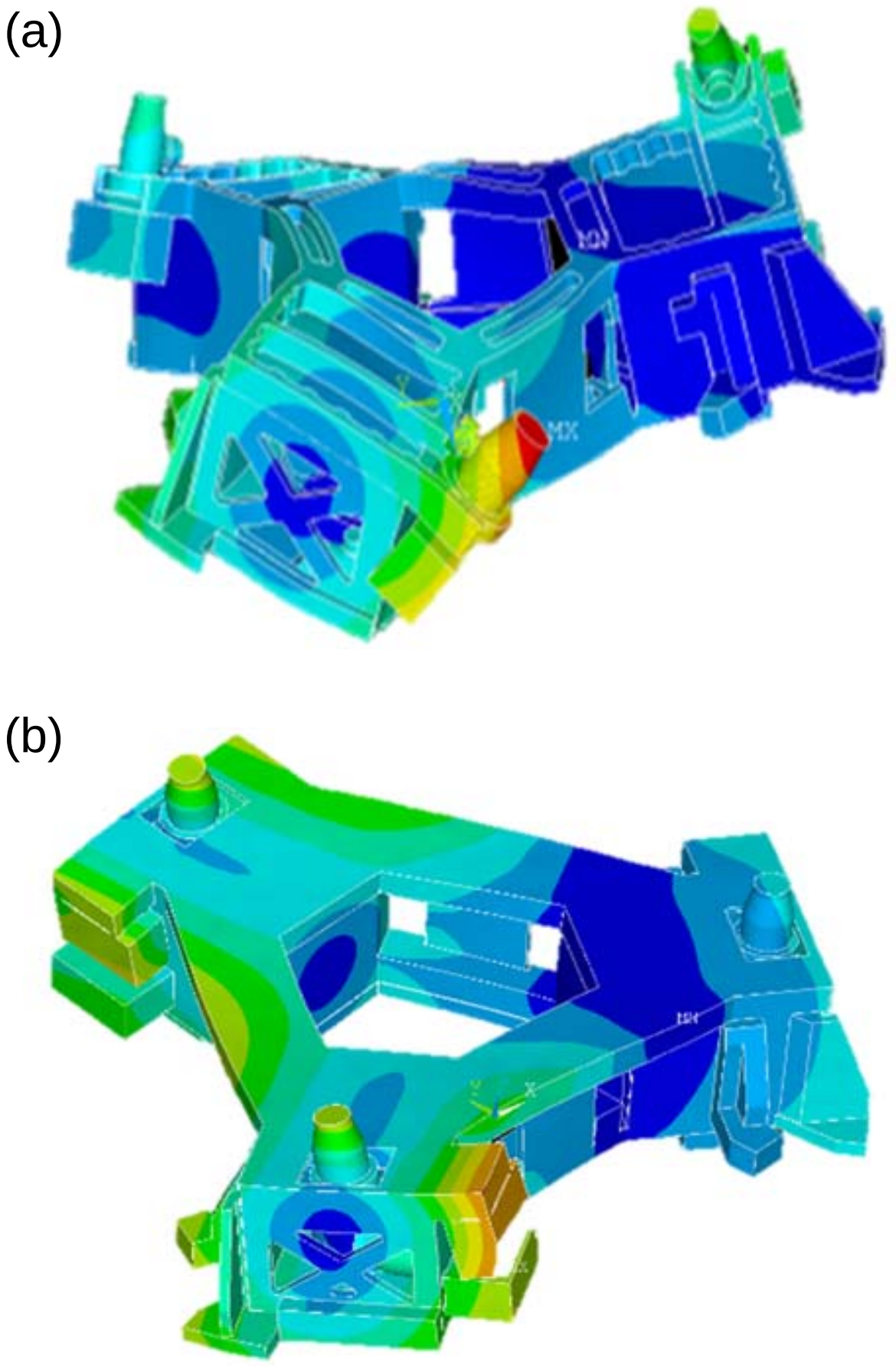

**Fig. 5. FEA modal analysis results for Stage 1. (a) Prototype, lowest mode at 150 Hz. (b) Advanced LIGO design, lowest mode at 255 Hz.**

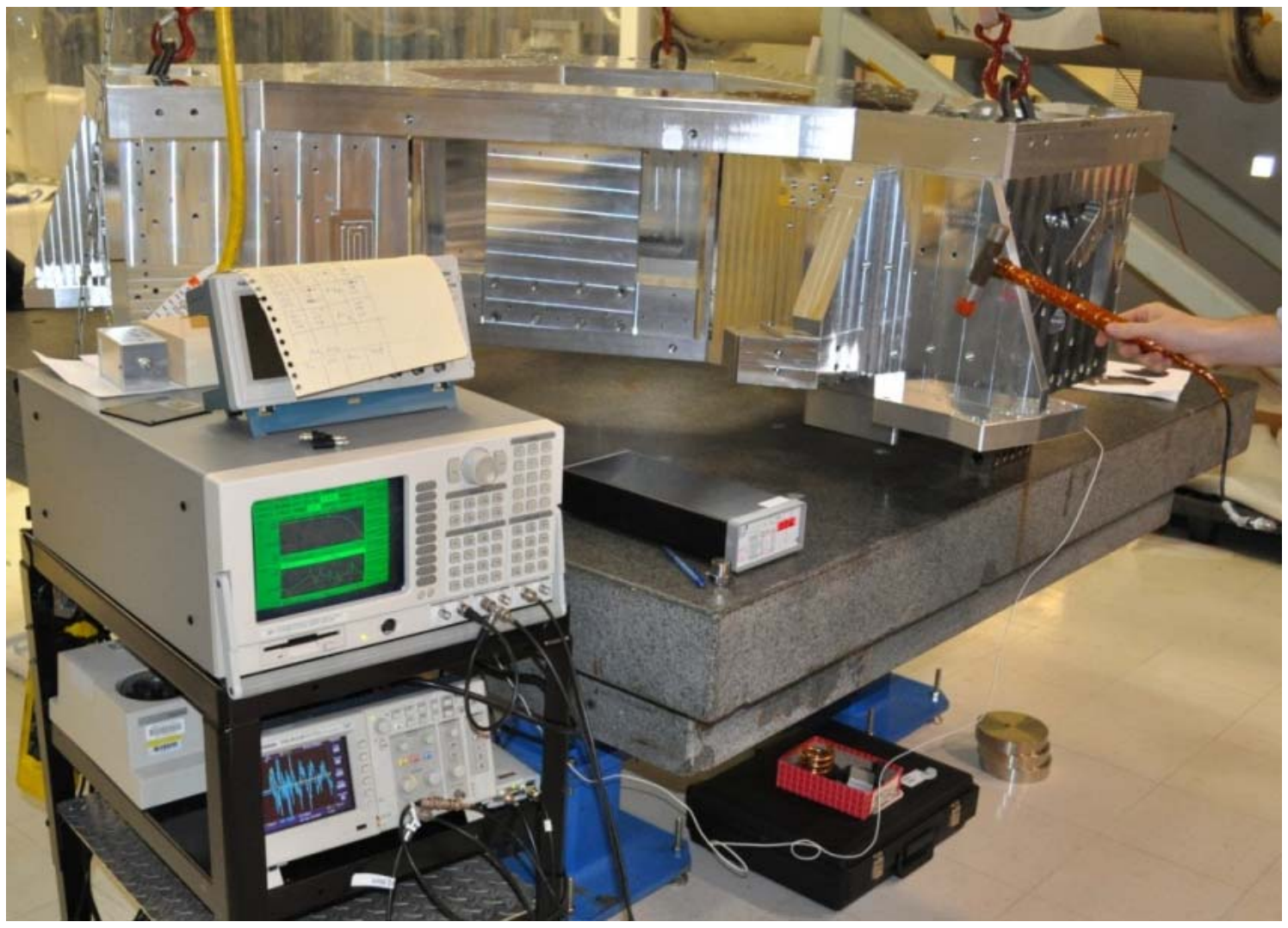

**Fig. 6. Stage 1 Modal Testing setup.**

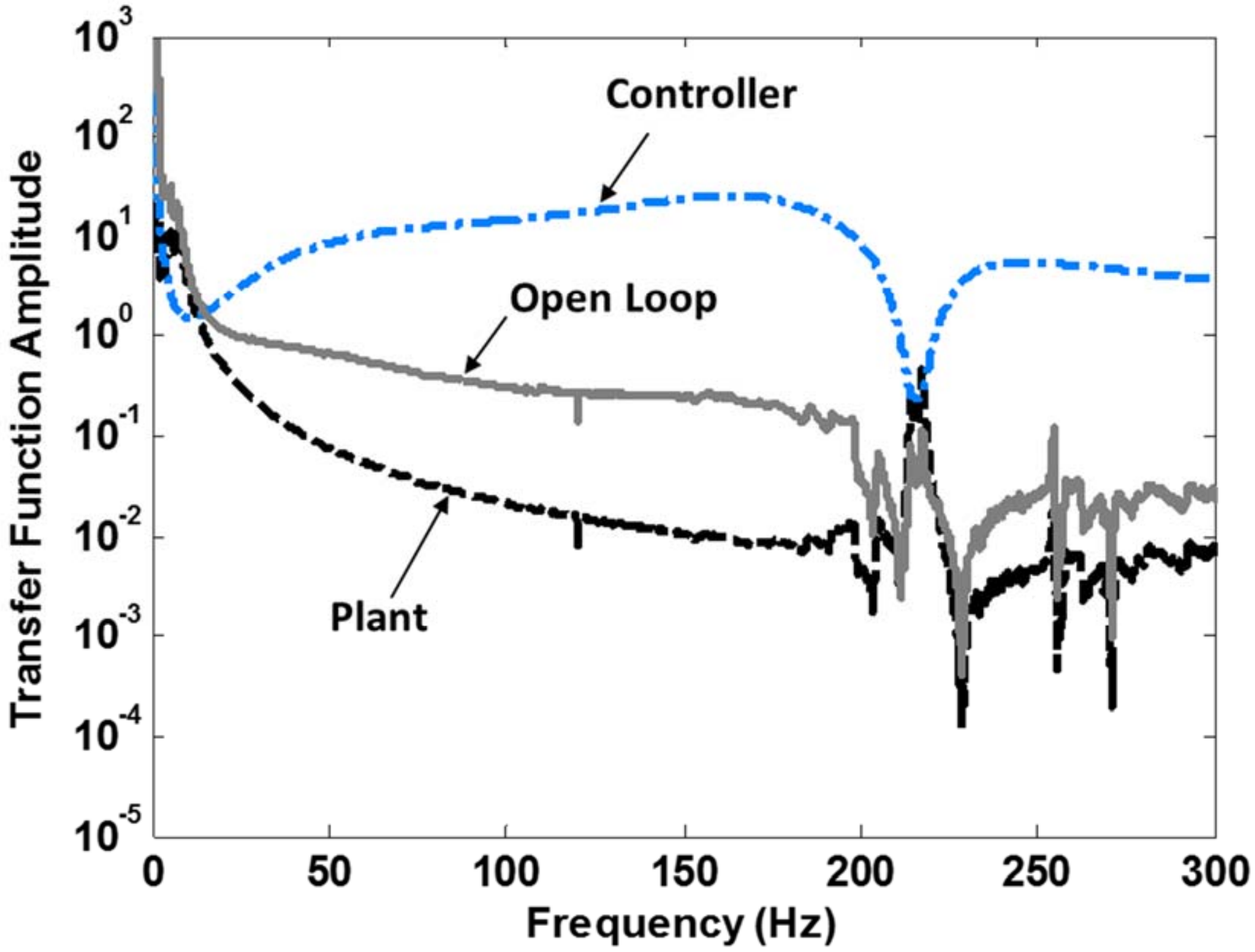

**Fig. 7. Advanced LIGO plant, controller and open loop transfer function.**

### 3.2 Equipment

The active control performance can be significantly affected by the couplings between the platform and its payload (equipment). Fig. 8 (a) shows the external frame of one of the Advanced LIGO payloads. It is a quadruple pendulum used to provide passive isolation to the interferometer optics. Analytical and experimental modal analysis were carried out to identify the modal shapes. The photo in Fig. 8 (b) shows a modal characterization test being performed. The experimental modal shapes identified for the lowest frequency mode is shown in Fig. 8 (c). It is a flag mode of the quadruple pendulum frame. More details can be found in [38]. Experimental transfer functions showed that even very small components mounted on the optical table could couple strongly with the large and heavy structure on which they were attached.

Several options to damp the structure were investigated [39]. Mass dampers installed on the payload frame prove to be a simple and very effective solution. Fig. 9 (a) shows vibration absorbers mounted on the equipment’s frame (top bracket not installed). A conceptual representation of the mass damper is shown in Fig. 9 (b), and a picture of a unit is shown in Fig. 9 (c). The mass dampers are made of 4 kg stainless steel mass. Rubber pads made of Viton are used as a spring and dissipative material. This material was chosen for being ultra-high vacuum compatible and for its excellent dissipation properties. Fig. 10 shows the damping which was obtained after installing passive damping components on the structure. The large resonance near 100 Hz has been reduced by more than a factor of 50. Best results are obtained when the Viton pads are the least pre-loaded, as shown in Fig. 9 (a) (no top bracket). When installing the top bracket of the vibration absorber, the tension in the assembly must be well controlled, as illustrated in Fig. 9 (b), to not compromise the damping effect.

These passive damping results significantly simplified the control commissioning and improved the system robustness. The technique has been generalized to damp either global or local modes. All the Advanced LIGO suspension frames have been equipped with mass dampers, and the BSC-ISI ballast masses are mounted on Viton pads to help damping the internal modes of the platform.

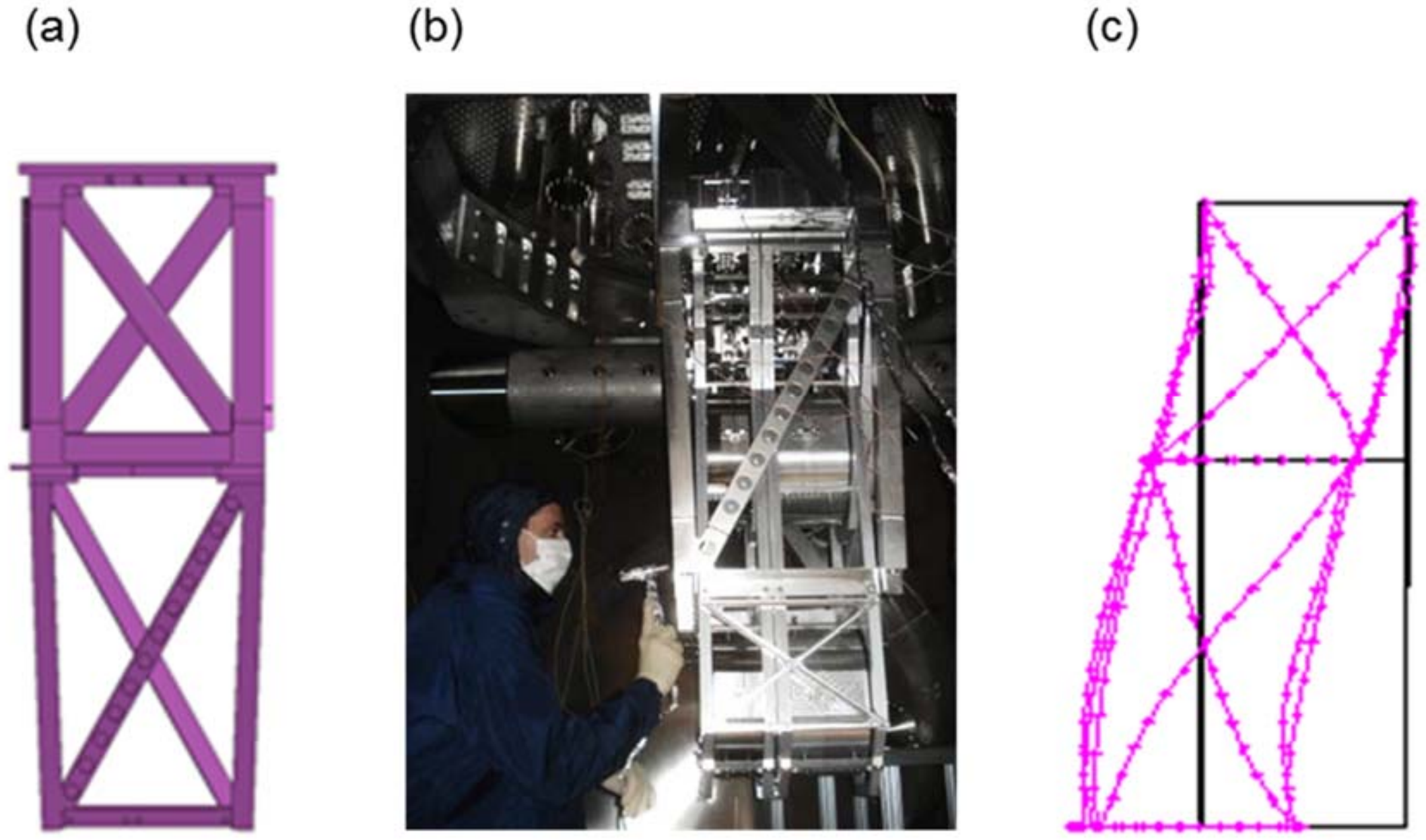

**Fig. 8. Optical payload (equipment) mounted on Stage 2 of the BSC-ISI. (a) CAD representation of the equipment external structure. (b) Modal testing of the equipment attached to the BSC-ISI. (c) Flag mode of the equipment at 81 Hz.**

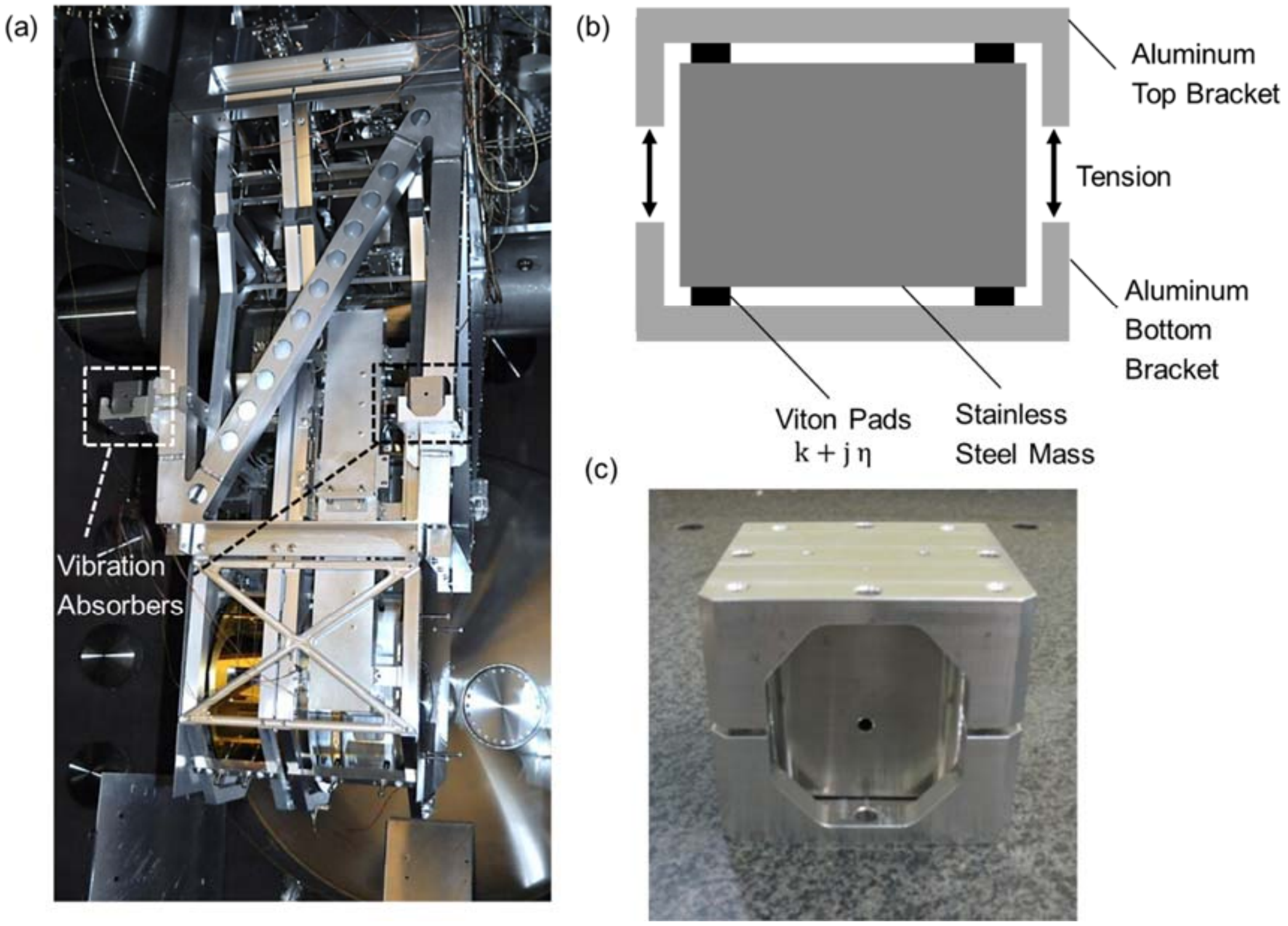

**Fig. 9. Vibration absorbers designed to damp the equipment resonances. (a) Vibration absorbers installed on the equipment's structure. (b) Conceptual representation of the vibration absorber. (c) A vibration absorber unit.**

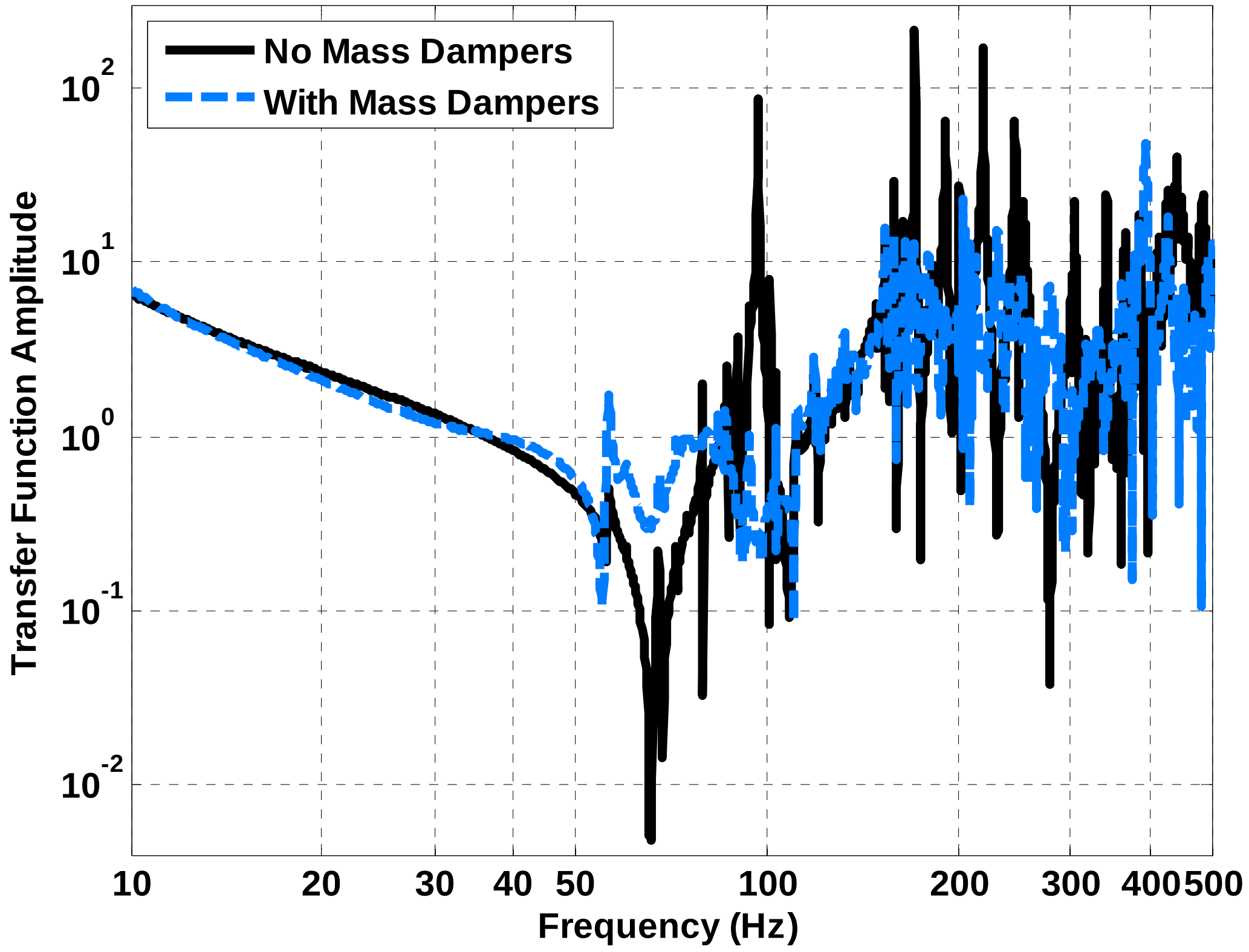

**Fig. 10. Transfer function of Stage 2 of the prototype with and without passive dampers on the equipment.**

## 4 Driven response

This section presents the force driven response of the system, both for translational and rotation degrees of freedom. The goal is to show that this 12 degrees of freedom platform behave as a two-mass spring system in each Cartesian direction as intended by design.

The curves in Fig. 11 and Fig. 12 show transfer functions from a force (or torque) applied on Stage 1 to the motion of Stage 2. In these measurements, the active inertial damping is used damp the rigid-body mode resonances. The transfer functions are normalized by the Stage 0-1 spring Stiffness so that the DC response is equal to unity.

Fig. 11 shows the response of the pitch and vertical degrees of freedom. The dashed curve shows the transfer function from a torque applied on Stage 1 along the pitch axis, to the rotation motion of Stage 2 around the same axis. The response along the roll axis (not shown) is similar to the response along the pitch axis. The solid curve shows the transfer function from a force applied along the vertical axis, to the translation motion along the same axis. Above the second frequency mode, the slope of the curves is function of the fourth power of frequency. Both curves show near -40 dB of magnitude at 10 Hz and are under -100 dB of magnitude at 100 Hz.

The second plot shows the responses in the longitudinal and yaw directions. The dashed curve shows the transfer function from a torque applied along the yaw axis, to the rotation motion around the same axis. The solid curve shows the transfer function from a force applied along the longitudinal axis, to the translation motion along the same axis. As for the previous curves, the slope above the frequency mode is function of the fourth power of frequency. Both curves are under -40 dB of magnitude at 10 Hz, and under -120 dB of magnitude at 100 Hz.

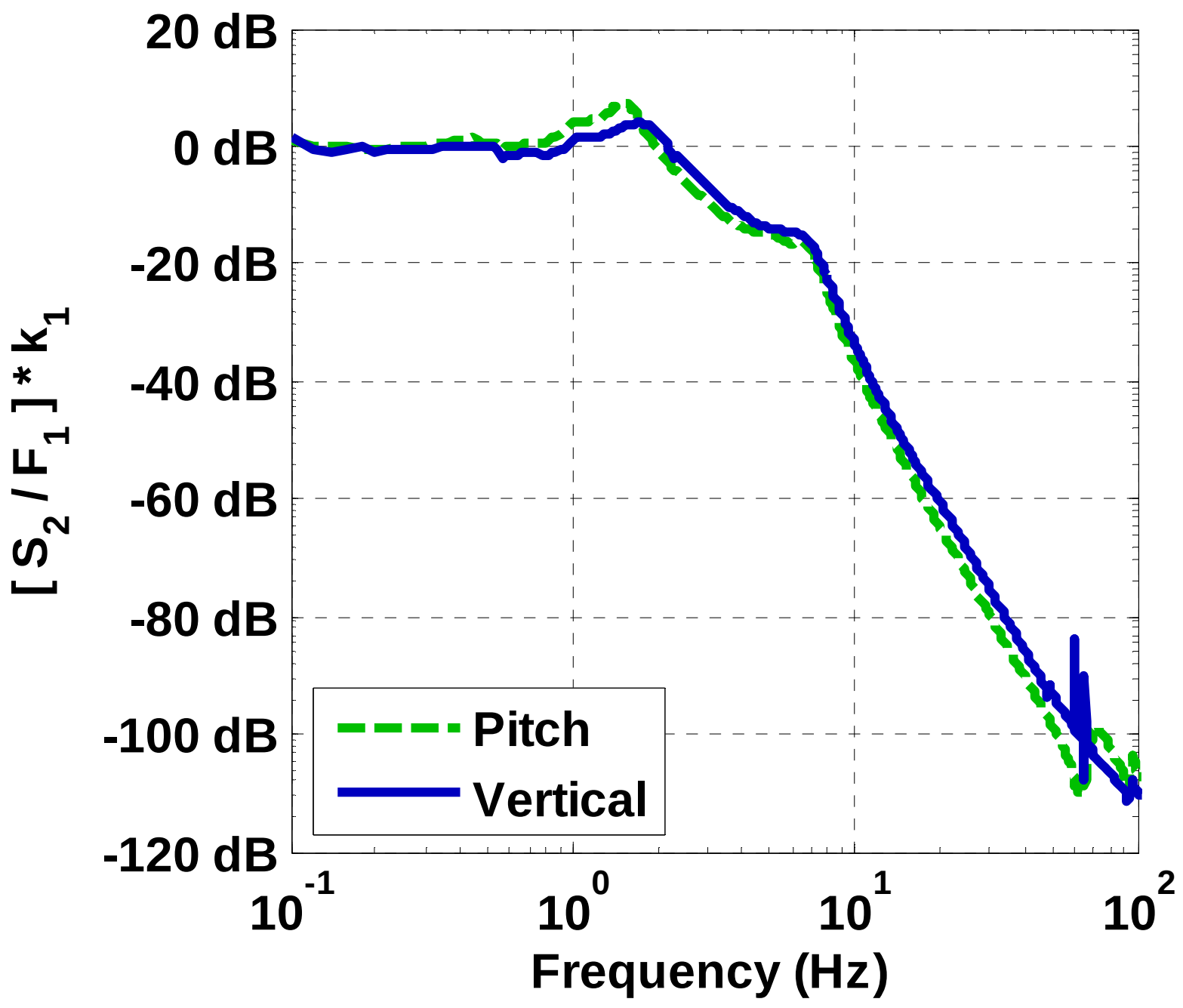

**Fig. 11 Stage 1 to Stage 2 driven transfer functions for the pitch and vertical degrees of freedom.**

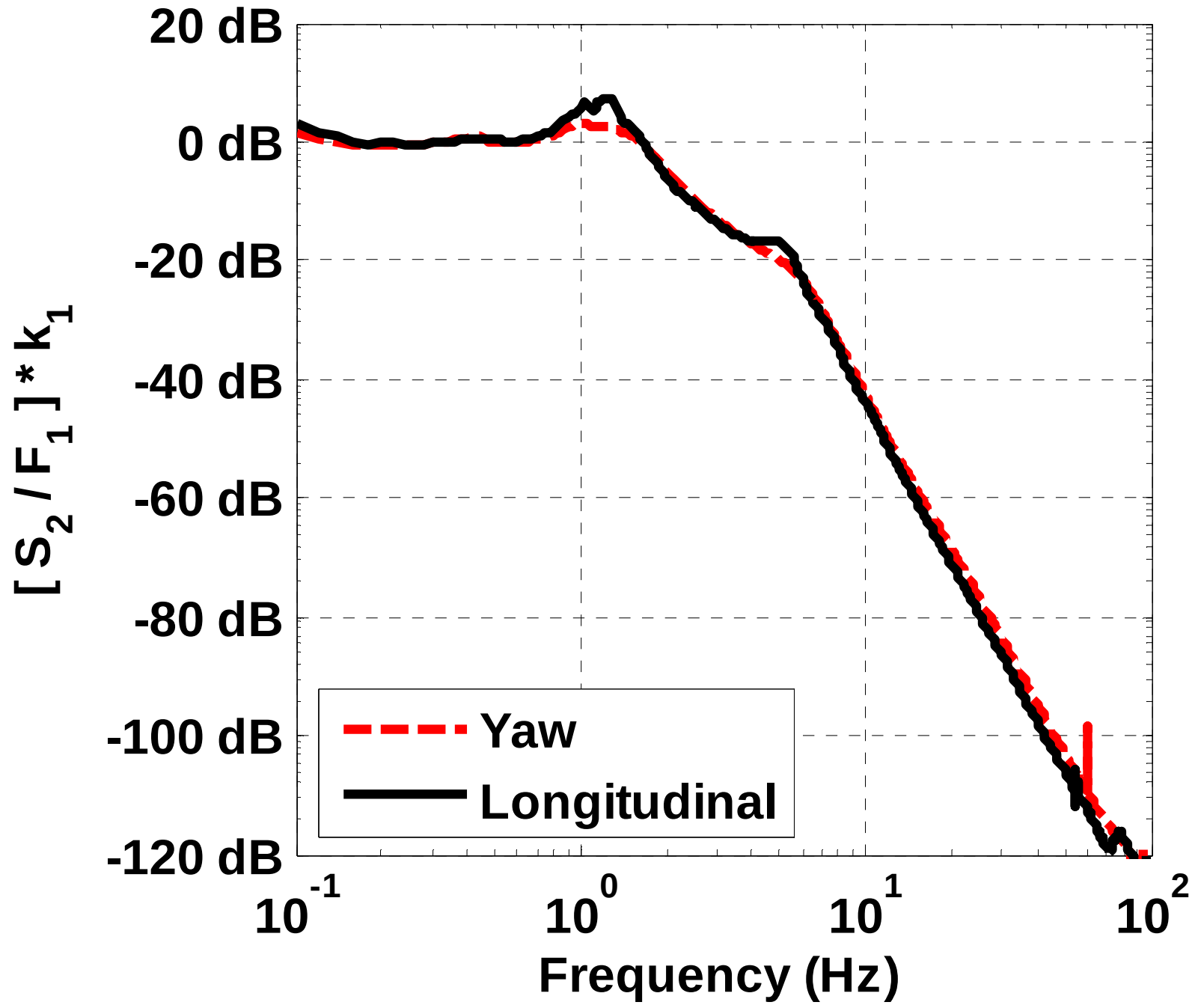

**Fig. 12 Stage 1 to Stage 2 driven transfer functions for the yaw and longitudinal degrees of freedom.**

## 5 Control Loops

This section summarizes the control strategy of the BSC-ISI system, and presents examples of feedback control loops. The core of the active isolation strategy is based on feedback control. All of the twelve degrees of freedom are controlled independently. The block diagram in Fig. 13 shows the control topology for one degree of freedom of Stage 1.

In this diagram, the control of the longitudinal motion of Stage 1 ($X_1$) is used as an example. Stage 1 motion is disturbed by the ground motion ($X_0$) through the seismic path (called $P_S$ in the equations) and controlled with the actuator force ($F$) through the force path (called $P_F$ in the equations). The absolute motion of Stage 1 motion is sensed with the geophones (*L4Cs*) and the 3 three-axis seismometers (*T240s*). The relative motion between the ground and stage 1 is measured with the six capacitive position sensors (*CPSs*). For each set of instruments, the individual signals are calibrated and combined to estimate the stage motion in the Cartesian basis (*Cart & Cal blocks*).

The signal from the L4C geophones is used to damp the rigid mode resonances with the damping filter $D$. This controller is a very robust velocity feedback loop that is engaged by default during the phases of testing and open-loop characterization. It reduces the risk of saturation at the resonances and reduces the dynamic range in order to ease and speed up the commissioning of the isolation loops.

The *CPS*, *T240* and *L4C* signals are combined in a sensor fusion using the low-pass filter $L$, the band-pass filter $B$, and high-pass filter $H$. At low frequencies, the filter $L$ passes the *CPS* signal to provide positioning capability. At higher frequencies, typically above 0.1 Hz, it filters the *CPS* signal to allow seismic isolation. The filters $B$ and $L$ combine the *T240* and *L4C* signals to provide a very low noise and broadband inertial sensing combination. At high frequencies (above 0.1 Hz), they pass the inertial sensing signal. At low frequency, they are designed to filter the noise of the inertial sensors. The signal resulting from this sensor fusion is sent to the feedback control filter $C_{FB}$ which is typically designed to obtain a unity gain frequency between 30 Hz to 40 Hz, and to provide high loop gain at low frequencies.

The sensor fusion filters are designed to be complementary as shown in Eq. (1), in order to facilitate the controller design and the performance analysis. Under those conditions, the closed loop response reduces to the expression given in Eq. (2) (it assumes that the damping filter effect $D$ is negligible when the control filter $C_{FB}$ is engaged). The noise term related to inertial sensing and the noise term related to relative motion sensing are introduced in the power spectra in Eq. (3), assuming that all the noise terms are uncorrelated. $N_{cps}$ is amplitude spectral density (ASD) of the capacitive sensor noise, $N_{T240}$ is ASD of the the T240 seismometer noise, and $N_{L4C}$ is the ASD of the L4C geophone noise.

In the control bandwidth, where the loop gain is high, the amplitude spectral density of the stage motion tends to the expression given in Eq. (4). This approximation can be used to design the fusion filters in order to minimize the motion as a function of the input motion and the sensor noise estimates.

$$H + \mathrm{B} + L = 1 \tag{1}$$

$$\frac{X_1}{X_0} = \frac{P_s - L\, C_{FB}\ P_F}{1 + C_{FB}\ P_F} \tag{2}$$

$$\begin{aligned} {X_1}^2 = {} & \left(\frac{P_s + L\, C_{FB}\ P_F}{1 + C_{FB}\ P_F}\right)^2 {X_0}^2 + \left(\frac{L\, C_{FB}\ P_F}{1 + C_{FB}\ P_F}\right)^2 {N_{CPS}}^2 \\ & + \left(\frac{B\ C_{FB}\ P_F}{1 + C_{FB}\ P_F}\right)^2 {N_{T240}}^2 + \left(\frac{H\ C_{FB}\ P_F}{1 + C_{FB}\ P_F}\right)^2 {N_{L4C}}^2 \end{aligned} \tag{3}$$

$$\lim_{(C_{FB}\ P_F) \to \infty} X_1 = \sqrt{(L\, X_0)^2 + (L\ N_{cps})^2 + (B\ N_{T240})^2 + (H\ N_{L4C})^2} \tag{4}$$

The same strategy is used for the feedback control of all other degrees of freedom, though the fusion filters are tuned differently for each of them. The tuning is done to optimize the motion of each degrees of freedom with respect to the input motion and the sensor noise. Special care is taken with the tuning of the pitch and roll degrees of freedom as low frequency motion amplification in those directions translates into unwanted signal in the horizontal seismometers through tilt-horizontal coupling [40].

Once the feedback loops are engaged, feedforward control can be used to obtain further isolation if there is residual coherence between the witness sensors (installed on the pre-isolator and on the ground) and the target sensors (Stage 1 inertial sensors). For that, a set of ground seismometers is used in a feed forward scheme called sensor correction. The ground instruments are high-passed with the high-pass filter ($H_{sc}$) before being combined with the relative sensors measurements. This control path results in additional isolation in the 0.1 Hz to 1 Hz range. Finally, a set of geophones mounted on the pre-isolator (*Stage 0 L4Cs*) can be used in a standard feed forward path through the controller $C_{ff}$ to obtain additional isolation in the 5 Hz to 25 Hz range. Eq.(5) gives the closed loop response including the sensor correction. If there is perfect coherence between the witness and target sensors then Eq (6) shows the improvement in the isolation. The ideal feed forward controller can be calculated as given in Eq. (7). Useful information on feed forward techniques can be found in [41].

$$\frac{X_1}{X_0} = \frac{P_s + L\, C_{FB}\; P_F\,(1 - H_{SC})}{1 + C_{FB}\; P_F} \tag{5}$$

$$\lim_{(C_{FB}\, P_F)\to\infty} \left(\frac{X_1}{X_0}\right) = L(1 - H_{SC}) \tag{6}$$

$$C_{FF} = -\frac{P_s}{P_f} - L\, C_{FB}\; (1 - H_{SC}) \tag{7}$$

Fig. 14 shows the control topology for one degree of freedom of Stage 2. It is similar to the control scheme used for Stage 1, except that there is only one set of inertial sensors in the feedback loop, and that the feed forward and sensor correction loops use Stage 1 instruments instead of Stage 0 instruments.

An example of control filters for the longitudinal direction is presented in Fig. 15. The solid curve shows the plant transfer function $P_F$ (Displacement over force. Amplitude is normalized to unity). The dotted curve shows the feedback controller $C_{FB}$ . It is designed to provide high bandwidth (40 Hz), and therefore it includes a few high frequency features to maintain adequate gain margin. The dash-dotted line shows the open loop. It has 45 degrees of phase margin, and provide high loop gain in the control bandwidth (about 100 at 1 Hz). The dashed curve shows the closed response to the force disturbance. The high bandwidth objective results in a bit of gain-peaking near the unity gain frequency, which is an excellent compromise since the Advanced LIGO interferometer is very insensitive to motion of the platform at those frequencies (motion is filtered by the passively in the next stages of isolation). Similar control loops are designed for all other degrees of freedom of Stage 1 and Stage 2.

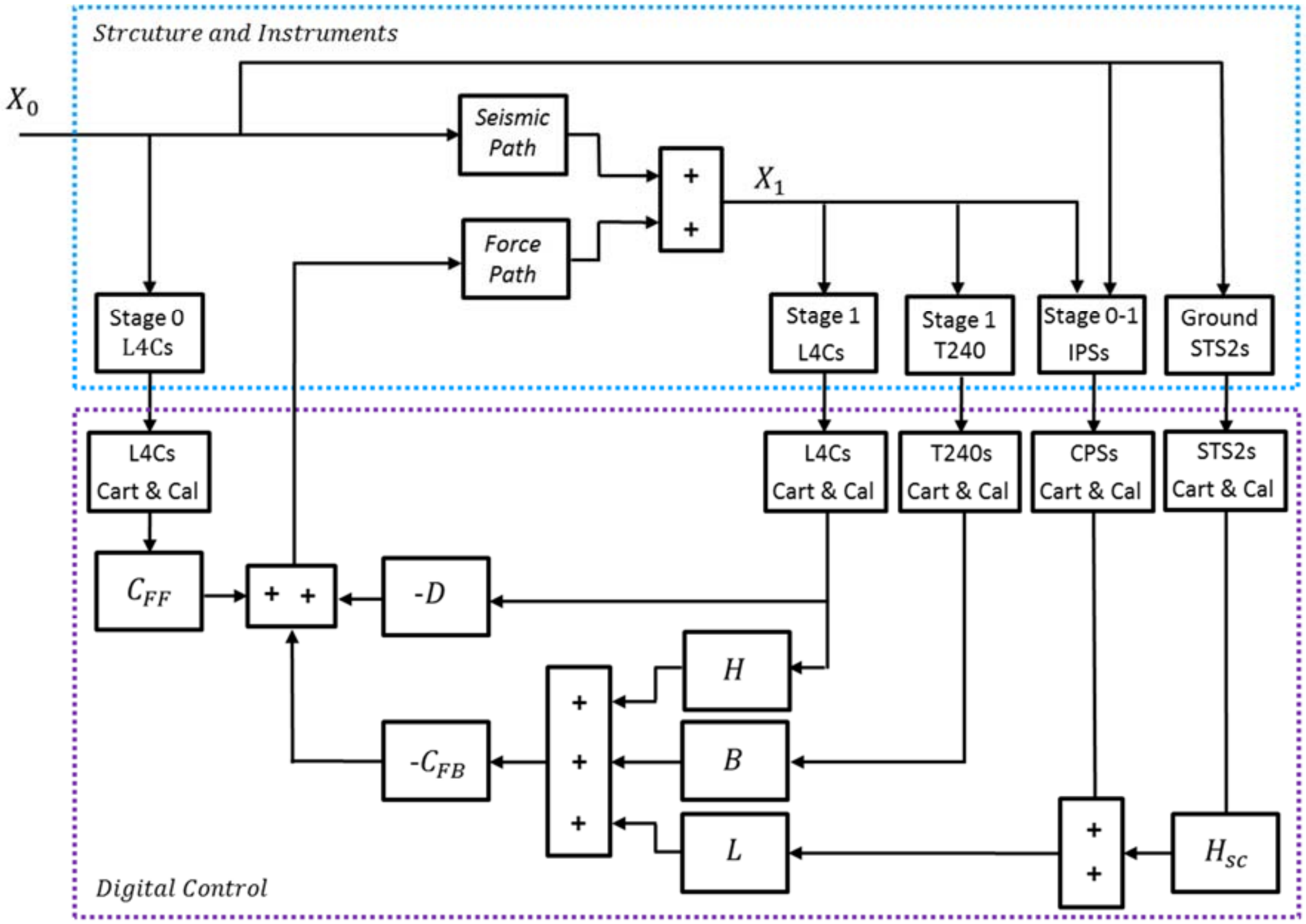

**Fig. 13. Control topology for Stage 1.**

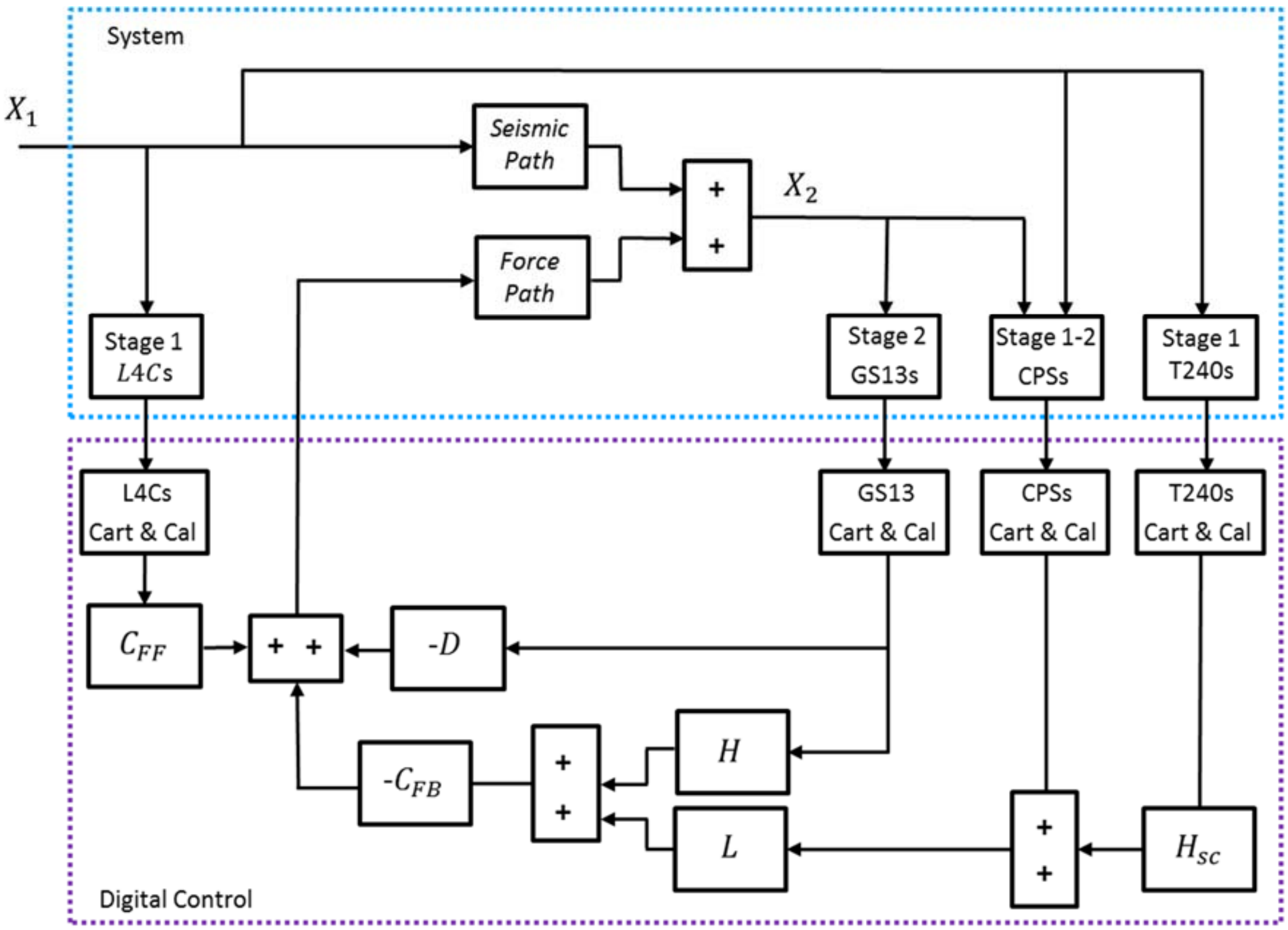

**Fig. 14. Control topology for Stage 2.**

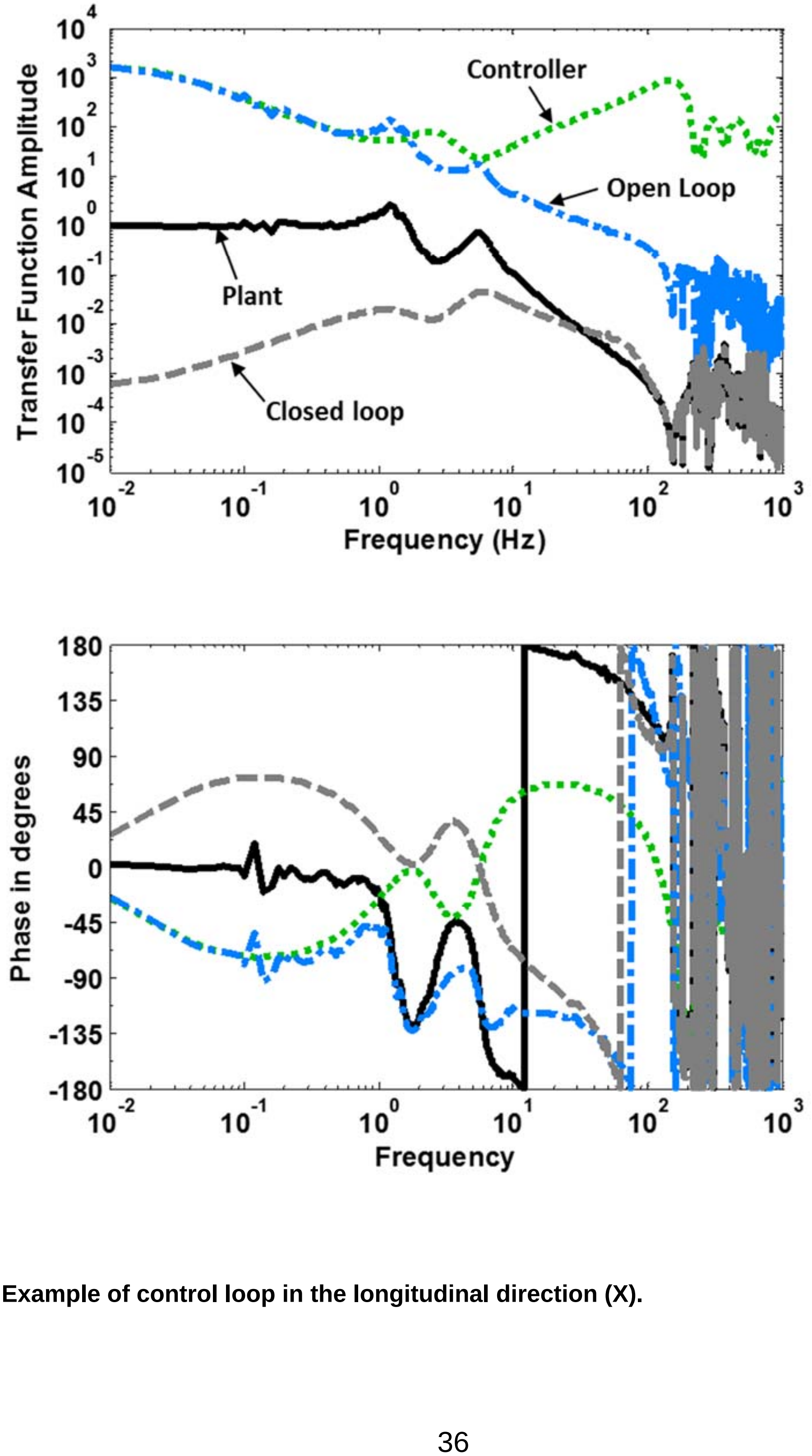

**Fig. 15. Example of control loop in the longitudinal direction (X).**

# 6 Isolation Tests

## *6.1 Transmissibility*

This section shows the seismic isolation provided by the system. All of the degrees of freedom are under control as described in section 5. To measure the system's transmissibility, the hydraulic actuators of the pre-isolator were used to drive Stage 0 motion. Geophones mounted on the pre-isolator are combined to estimate the input motion (Stage 0) in the Cartesian basis. The inertial sensors on Stage 2 are used to estimate the output motion along the direction of the drive. Transfer function measurements are performed in all directions of translation and rotation. Fig. 16 and Fig. 17 show the transmissibility up to 15 Hz. Above those frequencies the frame supporting Stage 0 deforms. Consequently, the sensors mounted on this frame do not provide an accurate measurement of the input motion.

In Fig. 16, the dashed curve shows transmissibility from Stage 0 to Stage 2 in the pitch direction. The controllers are tuned to provide approximately 20 dB of isolation at 1 Hz. Further isolation can be obtained at the cost of more noise injection at low frequency, which can add error in the horizontal inertial sensors' signal through tilt-horizontal coupling. The results obtained in the roll direction are similar to those obtained in the pitch direction.

The solid curve shows transmissibility from Stage 0 to Stage 2 in the vertical direction. For this direction, it is possible to tune the filters to provide much more isolation because the vertical motion does not affect the horizontal inertial sensors through tilt-horizontal couplings. In this example, the controller filters are tuned to provide about 45 dB of isolation at 0.4 Hz, and 70 dB at 1 Hz. These two specific frequencies correspond to payload natural frequencies at which the BSC-ISI must provide optimal performance.

In Fig. 17, the dashed curve shows transmissibility from Stage 0 to Stage 2 in the yaw direction. The same controllers are used as for pitch and roll. Further isolation can be obtained for this degree of freedom but it is often not necessary (ground yaw motion is typically small, and sensor signal is often close to sensor noise). Further noise analysis is currently being done to tune these filters.

The solid curve shows transmissibility from Stage 0 to Stage 2 in the longitudinal direction. The low frequency performance achievable in this direction (and in transversal) is limited by tilt horizontal coupling. At low frequency (around 100 mHz and below), the signal is dominated by tilt rather than horizontal motion [40]. In this example, the filters are tuned to provide 55 dB of attenuation at 1 Hz.

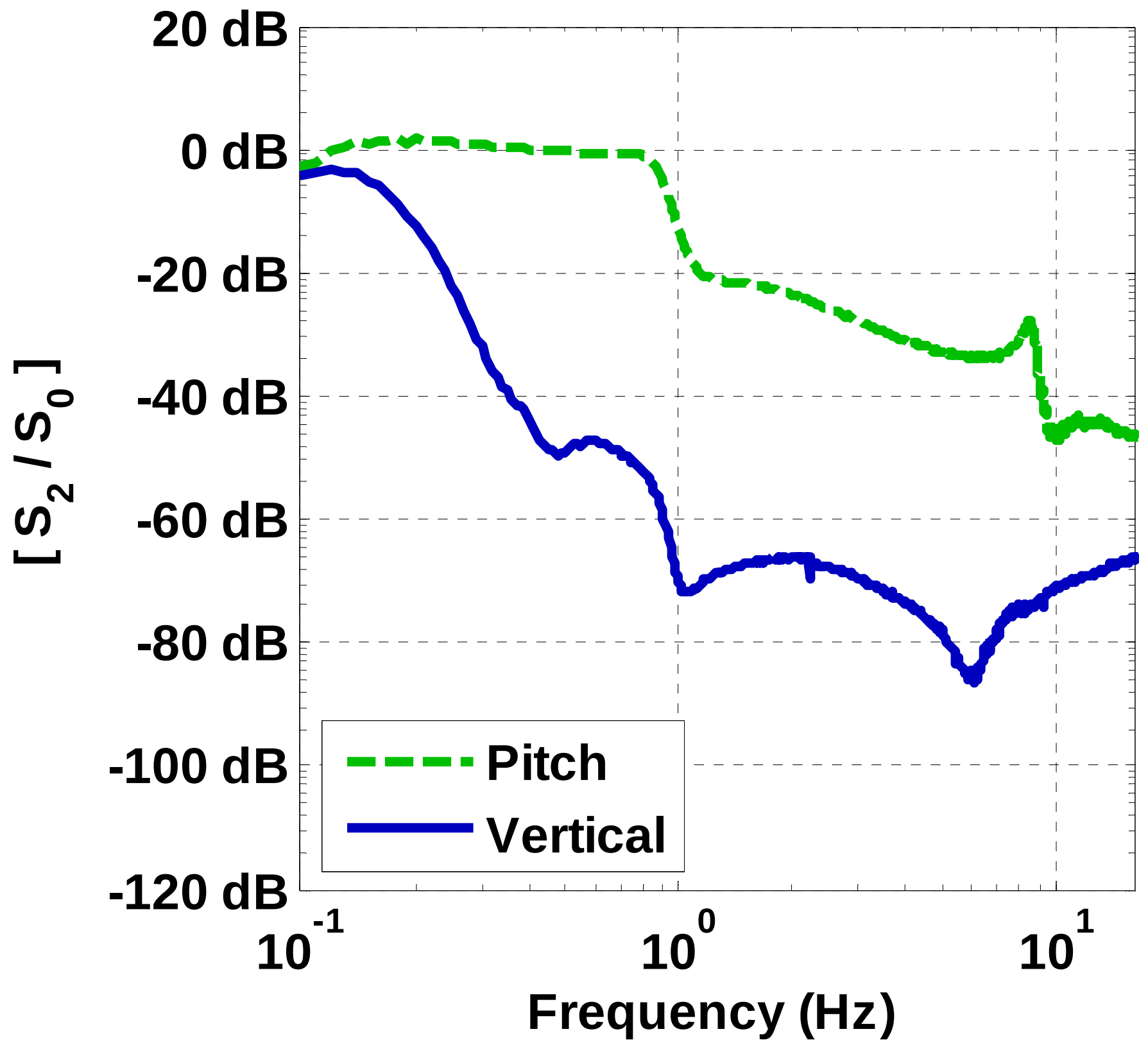

**Fig. 16. Vertical and pitch transmissibility.**

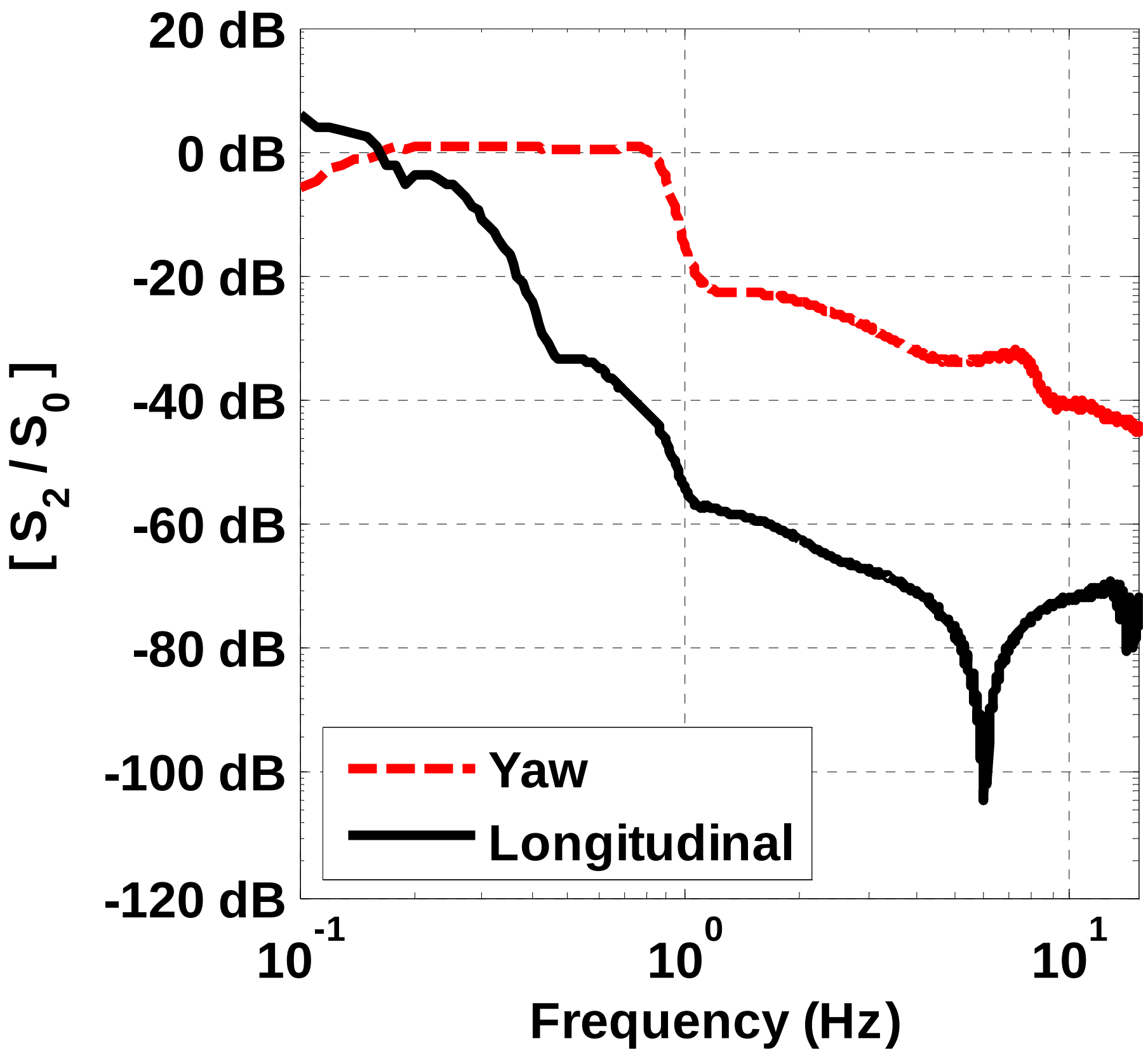

**Fig. 17. Longitudinal and yaw transmissibility.**

### *6.2 Absolute Motion Measurement*

This section presents the platform's absolute motion while it is actively controlled as described in section 5. A Streickeisen STS-2 is used to estimate the translational absolute ground motion (the experiment does not include a ground inertial rotation sensor). The inertial sensors on Stage 2 are used to estimate the rigid body motion of the platform's output.

Fig. 18 and Fig. 19 show the horizontal and vertical amplitude spectral density of the motion. In these two plots, the ground motion is shown by the solid line, the Advanced LIGO requirements are shown by the dash-dotted line, the inertial sensor theoretical noise is shown by the dotted line, the platform's motion measurement is shown in by the dashed line.

Up to 15 Hz, the platform motion is at or below the requirements. Above 15 Hz, the platform motion is very close to requirements. The small mismatch with requirements is inconsequential since seismic motion will not dominate the interferometer noise at those frequencies (the initial requirements included sufficient margin for such mismatch).

In the mid-band frequency [0.5 Hz to 10 Hz], the measurement is at or under the sensor noise. The portion of the curve under the sensor noise over-estimates the actual performance since those sensors are in loop. An out of loop witness sensor would be necessary to evaluate accurately the absolute motion in the frequency band. In-loop measurements under the theoretical sensor noise, however, indicate that there is room to sustain larger input motion and still maintain similar isolation performance. In this measurement done during the summer time at Hanford, the input motion was near $10^{-9}\, m/\sqrt{Hz}$ at 1 Hz. Measurements show that the motion at Livingston during the winter time can be more than 10 times larger. For such input, the output motion would still be near or slightly above the sensor noise. These results indicate that the BSC-ISI system should operate at or near requirements at most times.

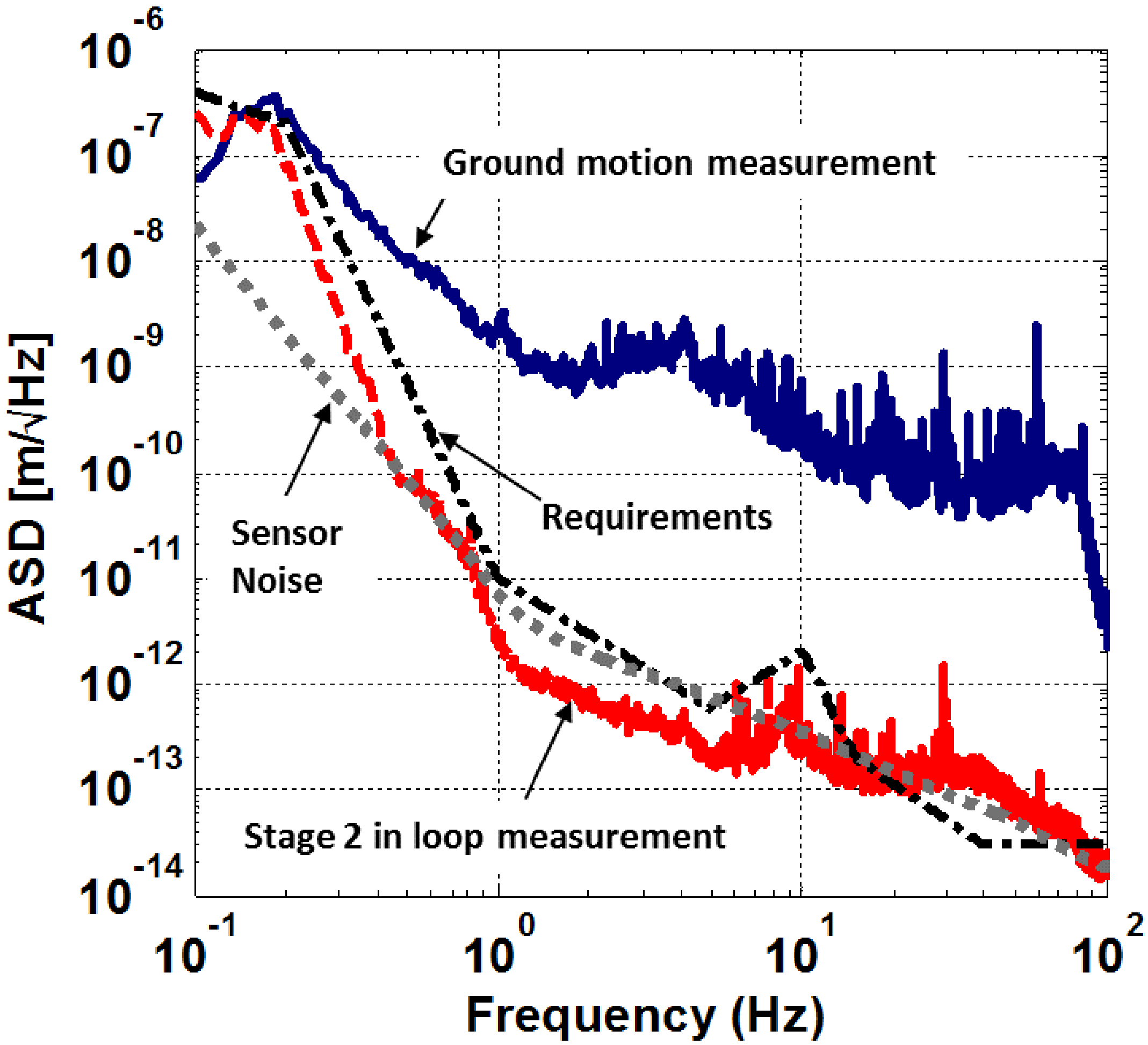

**Fig. 18. Amplitude spectral density showing the vertical seismic isolation (Z).**

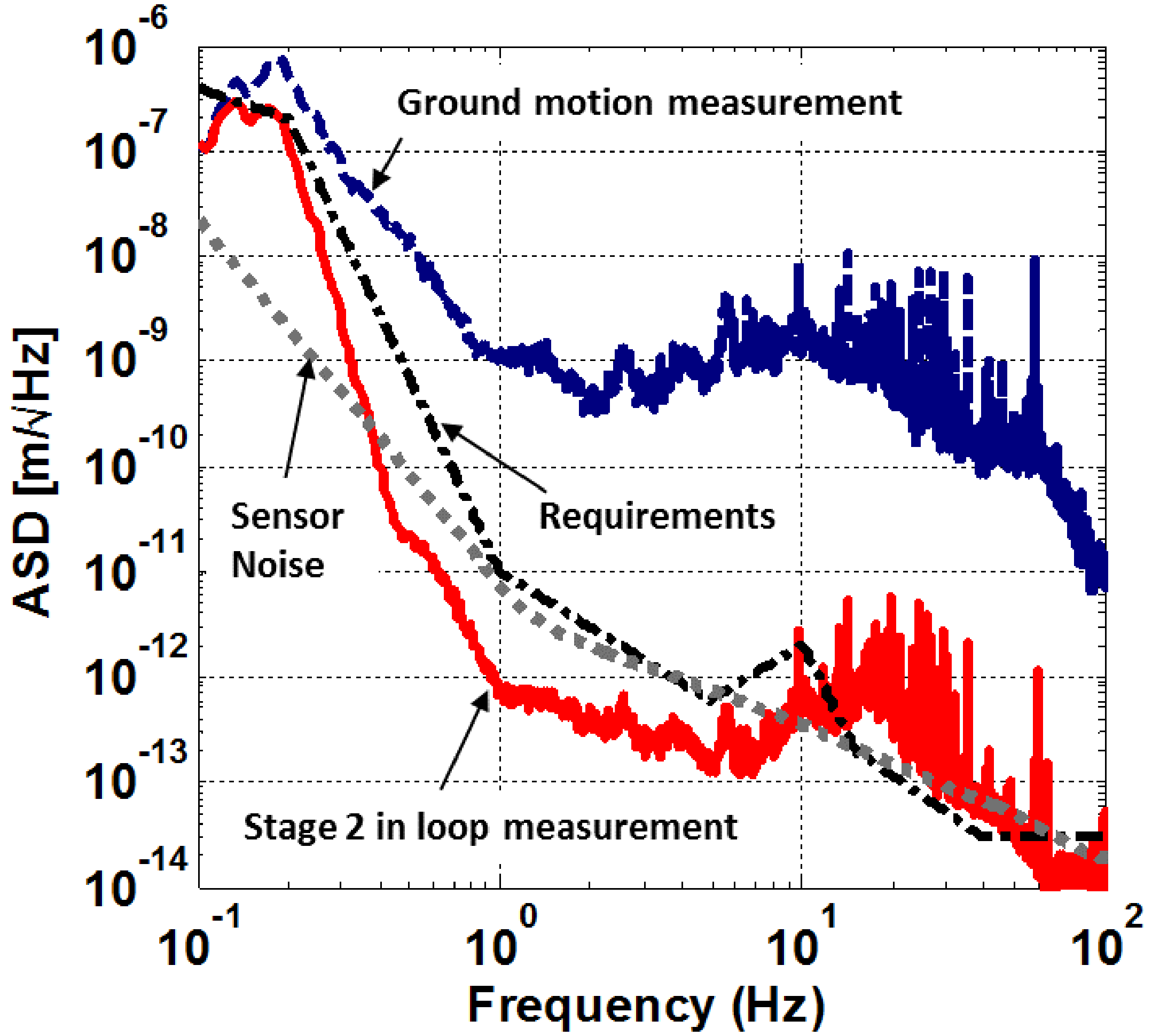

**Fig. 19. Amplitude spectral density showing the longitudinal seismic isolation (X).**

## 7 Conclusion

A prototype of a two-stage system designed for Advanced LIGO was built in 2006. This prototype has been tested and analyzed during the following two years. The results of this study led to the system's final design carried out in 2009 and 2010. The first unit was assembled and tested in 2011. Thirteen units have been built for the Advanced LIGO project during the past two years. The last two units are being constructed. The structural improvements done on the three stages of the final design allow the system to achieve a very high control bandwidth. The techniques implemented to passively damp the internal structural modes greatly improve the robustness of the feedback control. The engineering choices led to a very effective assembly and commissioning process. A BSC-ISI unit can be assembled and tested in less than four weeks. Experimental results have been presented. They show that the platform meets the very ambitious isolation requirements defined for Advanced LIGO more than a decade ago. In the coming years, the system's capability for tuning will be used to optimize the detector's performance at low frequencies. The platforms will support the operation of the interferometers on their way to detect the gravity waves predicted by Albert Einstein nearly a century ago.

## Acknowledgments

The authors acknowledge and gratefully thank the National Society Foundation for their support. LIGO was constructed by the California Institute of Technology and Massachusetts Institute of Technology with funding from the National Science Foundation and operates under cooperative agreement PHY-0107417.

We thank the JILA group for pioneering the work on active isolation systems using low frequency inertial sensors, and for demonstrating the feasibility of such multi-stage systems. We thank our colleagues from the suspension groups in GEO and LIGO for introducing us to the benefits of using triangular maraging steel blades to provide vertical isolation. We thank High Precision Devices for the mechanical design of the rapid prototype and the technical demonstrator. We thank Alliance Space Systems Incorporation for the mechanical design of the two-stage prototype. We thank Nanometrics, Streckeisen, Geotech, Sercel and Microsense for supplying us with great instruments, and for their technical support.

Finally yet importantly, this work would not have been possible without the outstanding support of the LIGO laboratory management, computer and data systems, procurement, facility modification and preparation, assembly and installation teams.

## List of Figures Captions

Fig. 1. CAD representation and picture of a BSC-ISI unit

Fig. 2. Schematic representation of BSC-ISI in the LIGO system

Fig. 3. CAD representation of BSC-ISI in the LIGO system

Fig. 4. Prototype transfer function and controller

Fig. 5. Comparison of prototype and final design FEA results

Fig. 6. Stage 1 Modal Testing

Fig. 7. Example of transfer function and control of the final design

Fig. 8. Optical payload (equipment) mounted on Stage 2 of the BSC-ISI

Fig. 9. Vibration absorbers designed to damp the equipment resonances

Fig. 10. Transfer function of Stage 2 prototype with and without passive dampers on the equipment

Fig. 11 Driven response for the vertical degrees of freedom

Fig. 12 Driven response for the horizontal degrees of freedom

Fig. 13. Control topology for Stage 1

Fig. 14. Control topology for Stage 2

Fig. 15. Example of control loops

Fig. 16. Vertical and Pitch Transmissibility

Fig. 17. Longitudinal and Yaw transmissibility

Fig. 18. Vertical Seismic Isolation

Fig. 19. Horizontal Seismic Isolation